\documentclass[pra]{revtex4-2}
\usepackage{braket}
\usepackage{graphicx}
\usepackage{color}

\begin{document}

\title{Origin of meter fluctuations in weak measurement interactions}

\author{Tomonori Matsushita} 
\email{d223728@hiroshima-u.ac.jp}
\author{Holger F. Hofmann} 
\email{hofmann@hiroshima-u.ac.jp}
\affiliation{Graduate School of Advanced Science and Engineering, Hiroshima University, Kagamiyama 1-3-1, Higashi Hiroshima 739-8530, Japan}

\begin{abstract}
Measurements map the value of a target observable onto a meter shift, resulting in a meter readout that combines the initial statistics of the meter state with the quantum statistics of the target observable. Even in the limit of weak measurement interactions, some information about the fluctuations of the target observable can be extracted from the change in the readout fluctuations caused by the measurement interaction. Here, we apply the Heisenberg picture to analyze the changes in the meter readout statistics caused by sufficiently weak measurement interactions, including the effects of non-linearities in the meter response. When additional information is obtained in a subsequent measurement of the system, the meter fluctuations are modified based on the post-selected statistics of the target observable. In addition, our analysis reveals a direct modification of the meter fluctuations due to the dependence of the post-selection probability on the dynamics induced by the meter in the measurement interaction. We point out that the quantum formalism makes it difficult to distinguish this dynamic term from the physical fluctuations of the target observable and stress the importance of distinguishing between genuine conditional fluctuations of the target observable and the dynamic pseudovariance associated with the measurement back-action.
\end{abstract}

\maketitle

\section{Introduction}
\label{sec:Introduction}
The goal of measurements is to extract information about a physical property of a system. The achievement of this goal requires an interaction between the system and an external meter so that we can access the information about the system property in this external meter. In classical physics, this was never considered to be a problem, since the effects of a classical force on a meter can always be directly proportional to the physical quantity that is being measured. The advent of quantum mechanics changed this situation in a fundamental manner, as Heisenberg famously explained in his introduction of the uncertainty principle \cite{Hei27}. Specifically, quantum mechanics introduces a trade-off relation between the disturbance of a system that is measured and the resolution of the value of the target observable in the measurement \cite{Mac06, Bus07, Wat11, Bus13, DreN14, Bcm14, Bus14, Roz15, Rod18, Mao19, Rod19, Pat19}. In a fully resolved measurement, the statistics of eigenvalues of the target observable are mapped onto the meter statistics, suggesting that the meter fluctuations limiting the resolution are completely independent of the statistics of the target observable. If that were the case in all measurements, it would be possible to identify the correct statistics of the target observable in measurements that are not fully resolved by removing the noise background associated with meter fluctuations using an appropriate deconvolution procedure. It should therefore be possible to extract information about the fluctuations of the target observable from the fluctuations of a meter after a measurement interaction that is too weak to resolve the eigenvalues of the observable. However, quantum theory indicates that the physics of such weak measurement cannot be understood in such a straightforward manner. As shown by Aharonov et al. \cite{Aha88}, the statistics of post-selected weak measurements cannot even be reconciled with the eigenvalues of the target observable. It is therefore necessary to consider the physics that map the value of a target observable onto the meter in more detail.

In the weak measurement limit, the meter fluctuations are dominated by the uncertainties of the initial meter state. The measurement interaction modifies the meter fluctuations only slightly, but the precise changes of the meter statistics can be identified with the corresponding statistics of the target observable responsible for the meter shift. The situation become more complicated when post-selection is performed on the system. As has been noted in previous research, post-selection can both increase and decrease the meter fluctuations, suggesting that the precision of the measurement depends on the post-selected result \cite{Par11, Par13, Ber15, Mat17}. On the other hand, Ogawa et al. have pointed out that changes of the readout fluctuations by post-selection can also be described by a "weak variance" derived entirely from complex weak values of the target observable and its square \cite{Oga21}. The reduction of the readout fluctuations by the post-selection is then explained by negative values of the "weak variance", and not by improvements of the precision of the measurement. Due to the anomaly of weak values, negative fluctuations of the target observable could be interpreted as a form of non-classical quantum correlation between post-selected results and the observable. However, it is also possible that the "weak variance" describes an update of the meter statistics provided by the changes of the post-selection probability due to the back-action dynamics. As we will show in the following, the changes of the meter fluctuations do indeed include a term that originates from this dynamical effect. Importantly, it is also necessary to consider the conditional uncertainties of the target observable, which can be done by distinguishing between the effect of forces physically changing the meter and the effect of information transfer resulting in a Bayesian update of information about a physical property before the interaction. These conditional fluctuations correspond to the statistical uncertainty of the post-selected weak values. It has been suggested that these fluctuations should be zero, corresponding to the precision associated with eigenvalues \cite{Vai17, Dzi19}. On the other hand, Hall \cite{Hal04} has argued that the measurement uncertainty defined by Ozawa in \cite{Oza03} can be identified with the statistical uncertainty of the weak values as the best estimate of a target observable when the post-selected outcome is identified with a measurement of that observable. According to this theory, the fluctuations of weak values are always positive and drop to zero for pure states with no imaginary part of the weak values \cite{Hof21,Lem22,Hof23}. The following analysis will show that the increase of the meter fluctuations caused by the conditional fluctuations of the target observable are indeed given by the Ozawa-Hall uncertainties, while the modifications that can result in a reduction of the meter fluctuations necessarily originate from additional information about the meter gained in the post-selection. Since the reduction of the meter fluctuations is independent of the intrinsic fluctuations of the target observable, this effect corresponds to a genuine modification of meter sensitivity by the post-selection, indicating that fluctuations in the meter sensitivity can be controlled by observing the effects of the back-action on the system.

In the present paper, we develop a general theory of weak measurement interactions based on derivatives of the Heisenberg equations of motion in the parameter that describes the strength of the measurement interaction. The Heisenberg equations of motion allow us to keep track of the physical changes of the meter that do not depend on the initial state of the meter or on any future post-selection of the system. Even without post-selection, the analysis shows that non-linearities in the system-meter interaction introduces modifications to the meter readout fluctuation. In particular, meter saturation effects can reduce the meter fluctuations without any improvement of meter precision. It is therefore important to consider the details of the meter dynamics when interpreting the fluctuations of the meter readout. Once the post-selection is included, the weak value replaces the expectation value in the average meter shift. Likewise, the Ozawa uncertainties replace the initial uncertainty of the target observable in the meter fluctuations. However, post-selection introduces an additional term in the meter fluctuations, and this term depends on the change of post-selection probability caused by the back-action of the meter on the system. Specifically, the post-selected outcome provides information about the magnitude of fluctuations in the generator of the meter dynamics and this information may change the meter readout fluctuation as a result of correlations between generator fluctuations and meter readout fluctuations in the initial meter state. We show that most meter states have a negative correlation between the two quantities, so that an indication of large back-action effects translates into a reduction of readout fluctuations, while an indication of small back-action effects result in increased readout fluctuations. We can thus show that a significant part of the changes to the readout fluctuations in a post-selected measurement originate from information about the meter before the interaction gained from the observation of the post-selected result. To evaluate the intrinsic conditional fluctuations of the target observable, these results of a Bayesian update of the meter statistics must be subtracted from experimentally observed results.

The rest of the paper is organized as follows. In Sec. \ref{sec:Heisenberg}, we introduce the Heisenberg equations of motion for the meter readout. In Sec. \ref{sec:Decomposition}, we analyze how the measurement interaction modifies the fluctuations of the meter readout and show that the changes of the meter fluctuations include non-linear effects of meter fluctuations and meter saturation. In Sec. \ref{sec:Cond.Hei.}, we introduce a subsequent measurement by including a projection operator in the Heisenberg picture dynamics and confirm that the weak values are obtained from the first derivative of the average readout. In Sec. \ref{sec:Cond.fluc.}, we analyze the effects of the subsequent measurement on the fluctuations of the meter readout. It is shown that the change of the post-selection probability by the back-action dynamics of the measurement interaction provides information on the initial meter fluctuations that can result in a conditional reduction of meter fluctuations. In Sec. \ref{sec:typical}, we apply the analyses in this paper into continuous variable meters and Gaussian meter states and compare our analysis with the results of \cite{Oga21}. In Sec. \ref{sec:discussion}, we discuss the possibility of evaluating the conditional fluctuations of the target observable from a measurement of the meter readout fluctuations. Sec. \ref{sec:Conclusion} summarizes the results and concludes the paper.

\section{Dynamics of information transfer in weak interactions}
\label{sec:Heisenberg}
A measurement of a physical property $\hat{A}$ of a system requires an interaction between the system and an external meter that depends on the specific value of the physical property. In general, the dependence of the meter response on the value of $\hat{A}$ creates entanglement between the system and the meter, correlating the value of $\hat{A}$ with the meter observable $\hat{M}$ so that a readout of $\hat{M}$ provides quantitative information on the value of $\hat{A}$. In sufficiently strong interactions, the statistics of $\hat{M}$ correspond closely to the statistics of the eigenvalues of $\hat{A}$ in the initial system state, and the fluctuations of the readout $\hat{M}$ can be explained mostly in terms of these input fluctuations of $\hat{A}$.  However, the precise readout statistics of $\hat{M}$ always contain fluctuations that originate from the initial meter state preparation and from the system-meter interaction. This problem becomes especially important when the measurement interaction is weak and the fluctuations of $\hat{M}$ in the initial meter state exceed the differences between the meter shifts associated with different eigenvalues of $\hat{A}$. In this case, it is difficult to distinguish the fluctuations associated with input uncertainties of $\hat{A}$ from the fluctuations of $\hat{M}$ that originate from the meter preparation and its effects on the interaction dynamics. In the following, we will therefore analyze in detail how the measurement interaction modifies the fluctuations of $\hat{M}$ in the limit of weak interactions. 

An ideal measurement interaction sensitive only to the physical property of the system $\hat{A}$ can be described by a bilinear Hamiltonian where the meter dynamics is generated by a meter operator $\hat{B}$. The interaction Hamiltonian is then proportional to a dyadic product of $\hat{A} \otimes \hat{B}$. It should be noted that the addition of any other terms to the interaction Hamiltonian would only introduce meter responses related to physical properties other than $\hat{A}$, resulting in unnecessary and avoidable measurement errors \cite{Mat21}. Since a measurement should be completed in a finite amount of time, the interaction should be time-dependent and its effect can be summarized by a unitary operation obtained from the time integral of the interaction Hamiltonian. The complete measurement interaction can then be expressed by
\begin{equation}
\label{eq:unitary}
\hat{U}_{SM}(s) = \exp \left(- \frac{i}{\hbar} s \hat{A} \otimes \hat{B}\right),
\end{equation}
where $s$ determines the effective strength of the interaction. The interaction causes a conditional transformation of the meter properties that depends on the target observable $\hat{A}$. We should therefore examine how the value of $\hat{A}$ appears in the meter readout represented by the operator $\hat{M}$. In the Heisenberg picture, this is expressed by the dependence of the readout operator $\hat{M}$ on the applied interaction strength $s$,
\begin{equation}
\label{eq:metershift} 
\hat{M}(s) = \hat{U}^\dagger_{SM}(s) \left(\hat{I} \otimes \hat{M}\right) \hat{U}_{SM}(s).
\end{equation}
Here, we use the dyadic product for the operators before the interaction to indicate that the initial states of the system and the meter are independent of each other. On the other hand, the $s$-dependent operators all exist in the collective Hilbert space of the meter and the system.  

As mentioned above, we are specifically interested in the effects of weak interactions characterized by a sufficiently small interaction strength $s$. In this case, the main part of the meter response can be characterized by the derivative of the dependence of the meter readout $\hat{M}(s)$ on the interaction strength $s$, 
\begin{equation}
\label{eq:lshift}
\left.\frac{d}{ds} \hat{M}(s)\right|_{s=0} = \hat{A} \otimes \hat{\Gamma}_M,
\end{equation}
where
\begin{equation}
\hat{\Gamma}_M = \frac{i}{\hbar} \left[\hat{B}, \hat{M}\right].
\end{equation}
According to Eq.(\ref{eq:lshift}), the meter shift is described by the dyadic product of a meter operator $\hat{\Gamma}_M$ and the physical property of the system $\hat{A}$. The operator $\hat{\Gamma}_M$ expresses the rate of change of the meter readout $\hat{M}$ in the internal meter dynamics given by the unitary operation
\begin{equation}
\label{eq:Munitary}
\hat{U}_{M}(\varphi_B) = \exp \left(-\frac{i}{\hbar} \varphi_B \hat{B}\right),
\end{equation}
where $\varphi_B$ is a parameter that quantifies the effects of the dynamics on the meter \cite{Mat21}. The operator $\hat{\Gamma}_M$ can then be defined as the derivative of the dependence of the meter readout $\hat{M}(\varphi_B) = \hat{U}^{\dagger}_{M}(\varphi_B) \hat{M} \hat{U}_{M}(\varphi_B)$ on the parameter $\varphi_B$,
\begin{equation}
\hat{\Gamma}_M = \left.\frac{d}{d\varphi_B} \hat{M}(\varphi_B)\right|_{\varphi_B=0}.
\end{equation}
In the following, we will refer to this operator as the meter response $\hat{\Gamma}_M$ since it quantifies the linear response of the meter to external forces. The meter response $\hat{\Gamma}_M$ therefore expresses the proportionality between the linear shift of $\hat{M}$ and the system property $\hat{A}$ as shown in Eq.(\ref{eq:lshift}). The expectation value of $\hat{\Gamma}_M$ before the interaction is determined by the initial meter state. The change in the expectation value of $\hat{M}$ caused by the measurement interaction can thus be given by
\begin{equation}
\left.\Braket{\frac{d}{ds} \hat{M}(s)}\right|_{s=0} = \braket{\hat{\Gamma}_M}\braket{\hat{A}}.
\end{equation}
For sufficiently small values of $s$, the expectation value of $\hat{A}$ is proportional to the expectation value of the meter readout $\hat{M}(s)$ after the interaction, describing the dynamics of a meter shift conditioned by the value of $\hat{A}$ in the system in the linear limit. 

By expressing the meter readouts in the Heisenberg picture, it is possible to describe the complete quantum statistics of the readout in terms of the initial conditions given by the preparation of the meter state before the measurement. It is easy to see that the initial fluctuations of $\hat{M}$ before the measurement interaction will contribute to the fluctuations of the meter readout after the interaction. However, a precise analysis will have to take into account the precise evolution of the meter fluctuations described by Eq.(\ref{eq:metershift}). In the following, we will therefore apply the Heisenberg picture analysis to the evolution of readout noise in the measurement interaction.

\section{Decomposition of meter noise}
\label{sec:Decomposition}
The Heisenberg picture can also describe how the measurement interaction changes the fluctuations of the meter readout $\hat{M}$. The variance of the meter readout after a measurement interaction of strength $s$ is given by
\begin{equation}
\label{eq:noisedef}
\Delta M^2(s) = \Braket{\hat{U}^\dagger_{SM}(s) \left(\hat{I} \otimes \hat{M}^2\right) \hat{U}_{SM}(s)} - \Braket{\hat{U}^\dagger_{SM}(s) \left(\hat{I} \otimes \hat{M}\right) \hat{U}_{SM}(s)}^2.
\end{equation}
At $s = 0$, the meter fluctuations correspond to the fluctuations of the meter readout $\hat{M}$ in the initial meter state $\ket{\phi}_M$. The interaction then adds fluctuations that might originate from a genuine uncertainty of the target observable $\hat{A}$ or from fluctuations in the meter dynamics, e.g. from fluctuations of the meter response $\hat{\Gamma}_M$. In the following, we are mostly interested in the limit of weak measurement interactions, so the goal will be to derive the lowest orders in $s$ that represent the changes in the meter fluctuations caused by the interaction. In general, it is possible to find a linear dependence of the meter fluctuations on $s$. However, this means that the response of the meter to positive values of $\hat{A}$ is fundamentally different from the response to negative values. If the meter is intended to be equally sensitive to positive and negative signals, it will necessarily satisfy a number of conditions that guarantee its symmetry. In the following, we will therefore consider the conditions that the meter has to satisfy in order to obtain a symmetric meter response. 

Since the meter readout $\hat{M}$ is proportional to the value of $\hat{A}$, a symmetric meter response requires that the eigenvalue spectrum of $\hat{M}$ is symmetric under an inversion of the sign of $\hat{M}$. For every eigenstate $\ket{m}$ with a positive eigenvalue $M_m$, there must be an eigenstate $\ket{-m}$ with a corresponding negative eigenvalue $-M_m$. The inversion can then be represented by a unitary transformation,  
\begin{equation}
\label{eq:inverse}
\hat{U}_{\mathrm{inv.}} \ket{m} = \ket{-m}.
\end{equation}
The statistical distribution of the meter readout $\hat{M}$ also depends on the initial meter state $\ket{\phi}_M$. In a symmetric meter setup, this state should be symmetric under the inversion $\hat{U}_{\mathrm{inv.}}$. Usually, one would chose a state with positive parity,
\begin{equation}
\label{eq:Ssymmetry}
\hat{U}_{\mathrm{inv.}} \ket{\phi}_M = \ket{\phi}_M.
\end{equation}
In addition to the conditions described by Eqs.(\ref{eq:inverse}) and (\ref{eq:Ssymmetry}), the dynamics of the meter shift should also be symmetric. Since the meter shift is generated by the meter operator $\hat{B}$, the application of $\hat{U}_{\mathrm{inv.}}$ inverts the direction of the meter response if
\begin{equation}
\label{eq:Bsymmetry}
\hat{U}^{\dagger}_{\mathrm{inv.}} \hat{B} \hat{U}_{\mathrm{inv.}} = -\hat{B}.
\end{equation}
If the conditions given by Eqs. (\ref{eq:inverse}) and (\ref{eq:Bsymmetry}) are satisfied, the inversion of the meter observable $\hat{M}$ after an interaction of strength $s$ is given by
\begin{equation}
\hat{U}^{\dagger}_{\mathrm{inv.}} \hat{M}(s) \hat{U}_{\mathrm{inv.}} = - \hat{M}(-s).
\end{equation}
This means that the inversion of the meter response can be represented by a negative interaction strength $-s$. For the quadratic terms that define the fluctuations $\Delta M^2(s)$, the inversions of the values of $\hat{M}$ cancel out and the meter fluctuations after an interaction with negative measurement strength are equal to the fluctuations obtained for a corresponding positive measurement strength,
\begin{equation}
\label{eq:Nsymmetry}
\Delta M^2(s) = \Delta M^2(-s).
\end{equation}
As expected, the symmetry conditions given by Eqs. (\ref{eq:inverse}), (\ref{eq:Ssymmetry}) and (\ref{eq:Bsymmetry}) guarantee that there cannot be any linear dependence of $\Delta M^2(s)$ on $s$ at $s=0$. It should be noted that these conditions will be satisfied by a wide variety of realistic meter systems, since there is no practical reason why a meter should be biased with regard to the sign of the input signal. For all unbiased meters, any effect of the interaction on the fluctuations of the meter readout can only appear in the second order dependence on measurement strength $s$.

In the Heisenberg picture, the second derivative of the operator $\hat{M}(s)$ is given by
\begin{equation}
\label{eq:nlshift}
\left.\frac{d^2}{ds^2} \hat{M}(s)\right|_{s=0} = \hat{A}^2 \otimes (-\hat{\Theta}_M),
\end{equation}
where 
\begin{equation}
- \hat{\Theta}_M = \frac{i}{\hbar} \left[\hat{B}, \hat{\Gamma}_M\right].
\end{equation}
Here, the operator $\hat{\Theta}_M$ expresses the reduction of the meter response $\hat{\Gamma}_M(\varphi_B)$ by the dynamics generated by $\hat{B}$. Similar to the meter response $\hat{\Gamma}_M(\varphi_B)$, this operator can be defined in terms of the dynamics described by the unitary transformation $\hat{U}_M(\varphi_B)$ given in Eq. (\ref{eq:Munitary}), 
\begin{equation}
\hat{\Theta}_M = - \left.\frac{d}{d\varphi_B} \hat{\Gamma}_M(\varphi_B)\right|_{\varphi_B=0}.
\end{equation}
In the following, we will refer to this operator as the saturation $\hat{\Theta}_M$ since it quantifies the decreases of the meter response $\hat{\Gamma}_M$ as the meter shift increases. It should be noted that the saturation is a non-linear effect of the meter dynamics. Since it is given by the second derivative of $\hat{M}(s)$, its effect on the interaction between the system and the meter will depend on the quadratic statistical moment of $\hat{A}$.

We can now analyze the second order dependence of $\Delta M^2(s)$ on $s$ by using Eqs.(\ref{eq:lshift}) and (\ref{eq:nlshift}). The second derivative is given by
\begin{equation}
\label{eq:decomposition}
\left.\frac{d^2}{ds^2} \Delta M^2(s) \right|_{s=0} = 2 \braket{\hat{\Gamma}_M}^2 \Delta A^2 + 2 \Delta \Gamma^2_M \braket{\hat{A}^2} - \braket{\hat{M} \hat{\Theta}_M + \hat{\Theta}_M \hat{M}} \braket{\hat{A}^2}.
\end{equation}
All of the terms that describe the changes in the readout fluctuations depend on the statistics of the target observable $\hat{A}$ in the input state of the system. However, only the first term corresponds to the correlation between meter shift and target observable expected from an ideal measurement. The second term depends on the fluctuations of the meter response $\hat{\Gamma}_M$ in the initial state of the meter. These fluctuations result in a variance of the meter shifts that is proportional to the magnitude of the target observable as given by the squared value of $\hat{A}$. Finally, the third term depends on the correlation of the saturation $\hat{\Theta}_M$ with the meter readout $\hat{M}$ in the initial meter state. This correlation indicates that the meter response will evolve differently depending on the actual value of $\hat{M}$. As the term ``saturation'' implies, this can mean that the increase of high values is smaller than the increase of low values, resulting in a compression of meter fluctuations. This is why the contribution of saturation to the change of meter fluctuations is negative in Eq.(\ref{eq:decomposition}). It should be noted that this reduction of readout fluctuations does not increase the precision of the measurement, since it is associates with a corresponding reduction of the average meter shift in higher derivatives in $s$. This example shows that it is not sufficient to analyze the magnitude of readout fluctuations when evaluating the precision of a measurement, pointing to a possible flaw in the conclusions reached in \cite{Ber15}. In the present case, a high value of $\hat{A}$ results in a lower-than-expected change of the meter readout $\hat{M}$, causing a reduction of the readout fluctuations by a term that is also proportional to the magnitude of the target observable given by the squared value of $\hat{A}$.

Eq.(\ref{eq:decomposition}) shows that the meter dynamics introduces additional changes to the readout fluctuations because of the dynamics of the meter response. The linear response of the meter to $\hat{A}$ is not constant because of a combination of initial fluctuations of the meter response $\hat{\Gamma}_M$ and the non-linearity of the meter response at higher values of $\hat{M}$. It should also be noted that the nonlinearity can actually reduce the amount of meter noise, so that the two sources of errors can compensate each other. In all possible cases, the increase in the meter fluctuations can be traced back to the statistics of $\hat{A}$ in the initial system state. Since the value of $\hat{A}$ is not changed by the interaction, there is no need to consider any possible effects of the back-action dynamics on the system. However, this situation changes when additional information about the system is obtained in a subsequent measurement. In the following, we therefore consider the representation of subsequent measurements and post-selection in the Heisenberg picture.

\section{Subsequent measurement after weak interactions} 
\label{sec:Cond.Hei.}
A subsequent measurement on the system after the measurement interaction updates our information concerning the value of $\hat{A}$. However, the outcome of the subsequent measurement also depends on the dynamics of the system associated with the value of the meter property $\hat{B}$. It is therefore important to distinguish the effects of additional information on the statistics of $\hat{A}$ from modifications of the meter statistics associated with the interaction dynamics. 

The outcome of a subsequent measurement can be expressed by an eigenstate of a physical property of the system $\ket{f}$. The statistics of the readout $\hat{M}(s)$ is conditioned by the outcome $\ket{f}$ of the subsequent measurement. In the Heisenberg picture, the selection of the outcome $\ket{f}$ is described by the projection operator $\ket{f}\bra{f}$ and the conditional meter readout after a measurement interaction of strength $s$ is given by
\begin{equation}
\label{eq:cond.shift}
\braket{\hat{M}(s \mid f)} = \frac{\left \langle\hat{U}^\dagger_{SM}(s) \left(\ket{f}\bra{f} \otimes \hat{M}\right) \hat{U}_{SM}(s)\right \rangle}{\left \langle\hat{U}^\dagger_{SM}(s) \left(\ket{f}\bra{f} \otimes \hat{I}\right) \hat{U}_{SM}(s)\right \rangle}.
\end{equation}
The dependence of the operators on $s$ is given by Eq.(\ref{eq:metershift}) for $\hat{M}_S$ and by
\begin{equation}
\label{eq:fchange}
\ket{f}\bra{f}(s) = \hat{U}^\dagger_{SM}(s) \left(\ket{f}\bra{f} \otimes \hat{I}\right) \hat{U}_{SM}(s)
\end{equation}
for the projector $\ket{f}\bra{f}$. The first derivative of the $s$-dependence of the projection operator is given by
\begin{equation}
\label{eq:lfchange}
\left.\frac{d}{ds} \ket{f}\bra{f} \right|_{s=0} = \frac{i}{\hbar} \left[\hat{A}, \ket{f}\bra{f}\right] \otimes \hat{B}.
\end{equation}
The commutation relation in this equation describes the system dynamics generated by $\hat{A}$,
\begin{equation}
 \left.\frac{d}{d\varphi_A} \ket{f}\bra{f}(\varphi_A)\right|_{\varphi_A=0} = \frac{i}{\hbar} \left[\hat{A}, \ket{f}\bra{f}\right],
\end{equation}
where $\varphi_A$ is the parameter of the unitary transformation describing the system dynamics only \cite{Mat21},
\begin{equation}
\label{eq:Sdynamics}
\hat{U}_{A}(\varphi_A) = \exp \left(-\frac{i}{\hbar} \varphi_A \hat{A}\right).
\end{equation}
The commutation relation thus expresses the linear response of the projector $\ket{f}\bra{f}$ to arbitrary external forces represented by $\varphi_A$, where Eq.(\ref{eq:lfchange}) shows the specific linear response to the back-action associated with the meter property $\hat{B}$. 

To determine the conditional meter readout, we also need to consider the dynamics of the product of $\hat{M}(s)$ and $\ket{f}\bra{f}(s)$. The first derivative of this product is given by 
\begin{equation}
\label{eq:lcond.out}
\left.\frac{d}{ds} \left(\ket{f}\bra{f}(s) \hat{M}(s)\right) \right|_{s=0} = \frac{1}{2} \left(\hat{A} \ket{f}\bra{f} + \ket{f}\bra{f} \hat{A}\right) \otimes \hat{\Gamma}_M + \left.\frac{d}{d\varphi_A} \ket{f}\bra{f}(\varphi_A)\right|_{\varphi_A=0} \otimes \frac{1}{2} \left(\hat{B} \hat{M} + \hat{M} \hat{B}\right).
\end{equation}
The first term of this equation represents the linear meter response to the value of $\hat{A}$ conditioned by the projector $\ket{f}\bra{f}$. Note that the correlation between $\hat{A}$ and $\ket{f}\bra{f}$ is represented by the anti-commutator of the two operators. The second term describes the back-action dynamics in the system conditioned by $\hat{B}$. Since we are looking at the derivative of the product of $\hat{M}(s)$ and $\ket{f}\bra{f}(s)$, the contribution of this term depends on the correlation between the meter generator $\hat{B}$ and the meter readout $\hat{M}$. Such a correlation exists when the meter fluctuations are biased towards one side for positive values of $\hat{B}$ and to the other for negative values of $\hat{B}$. In general, it is possible to find meter states that do not have such a conditional bias of their readout fluctuations. In fact, any correlation between $\hat{M}$ and $\hat{B}$ will result in an unnecessary increase in the uncertainty of $\hat{M}$, so it is natural to avoid any such correlations in the meter setup. In addition to the elementary conditions for the meter symmetry (\ref{eq:inverse}), (\ref{eq:Ssymmetry}) and (\ref{eq:Bsymmetry}), we therefore impose the additional requirement that the initial meter state $\ket{\phi}_M$ satisfies the condition 
\begin{equation}
\label{eq:MB}
\braket{\hat{B} \hat{M} + \hat{M} \hat{B}} = 0.
\end{equation}
We can then determine the first derivative of the conditional meter expectation value defined by Eq.(\ref{eq:cond.shift}),
\begin{equation}
\label{eq:cond.lshift}
\left.\frac{d}{ds} \braket{\hat{M}(s \mid f)} \right|_{s=0} = \braket{\hat{\Gamma}_M} \mathrm{Re}\left(\frac{\braket{f|\hat{A} \hat{\rho}|f}}{\braket{f|\hat{\rho}|f}}\right),
\end{equation}
where $\hat{\rho}$ is the density operator describing the input state of the system. The value of $\hat{A}$ that describes the meter shift conditioned by a subsequent measurement of $\ket{f}$ corresponds to the weak value of $\hat{A}$ associated with a pre-selection of $\hat{\rho}$ and a post-selection of $\ket{f}$. Our analysis thus confirms that weak values describe the outcomes of weak measurements up to the first derivative of the interaction strength $s$ for a wide variety of meter systems and meter state preparations.

The second term in Eq.(\ref{eq:lcond.out}) shows that the back-action dynamics in the system conditioned by the meter generator $\hat{B}$ has a non-trivial effect on the information obtained in the post-selection of a subsequent measurement outcome $\ket{f}$. We can therefore expect that the meter fluctuations $\Delta M^2(s)$ conditioned by $\ket{f}$ may include non-trivial effects of the meter dynamics associated with correlations between the readout $\hat{M}$ and the generator $\hat{B}$ responsible for the modifications of $\ket{f}\bra{f}$ by the back-action dynamics. It may be worth noting that a non-vanishing correlation between $\hat{M}$ and $\hat{B}$ means that the readout determines the value of the meter generator $\hat{B}$, corresponding to a measurement of the imaginary part of the weak value of $\hat{A}$ \cite{Joz07, Hof11, Dre12}. In Eq.(\ref{eq:lcond.out}), this contribution of the imaginary part of the weak value is represented by the derivative in the parameter $\varphi_A$. The first derivative of the conditional meter readout $\hat{M}(s|f)$ is therefore completely explained by the theory of weak measurements.

Since unbiased meter states will always satisfy the condition given by Eq. (\ref{eq:MB}), the effects of information about $\hat{B}$ gained in the post-selection of $\ket{f}$ do not contribute to the average meter shift and only the real part of the weak value determines this change of the meter state. However, there will be a non-vanishing contribution of this information gain caused by the back-action dynamics to the change of the meter fluctuations. The separation of these back-action related contributions from the Bayesian update of $\hat{A}$ based on the additional information provided by the outcome $\ket{f}$ will be the topic of the next section.

\section{Conditional meter fluctuations} 
\label{sec:Cond.fluc.}
As shown in the previous section, the Heisenberg picture can describe the effects of post-selection in terms of the dynamics of the projector $\ket{f}\bra{f}$ onto the post-selected outcome $\ket{f}$ in the system. We can now apply this formalism to examine how the information obtained in a subsequent measurement changes the fluctuations of the meter readout $\hat{M}$. The conditional fluctuations of the meter readout after a measurement interaction of strength $s$ are given by
\begin{equation}
\Delta M^2(s \mid f) = \frac{\left \langle\hat{U}^\dagger_{SM}(s) \left(\ket{f}\bra{f} \otimes \hat{M}^2\right) \hat{U}_{SM}(s)\right \rangle}{\left \langle\hat{U}^\dagger_{SM}(s) \left(\ket{f}\bra{f} \otimes \hat{I}\right) \hat{U}_{SM}(s)\right \rangle} - \left( \frac{\left \langle\hat{U}^\dagger_{SM}(s) \left(\ket{f}\bra{f} \otimes \hat{M}\right) \hat{U}_{SM}(s)\right \rangle}{\left \langle\hat{U}^\dagger_{SM}(s) \left(\ket{f}\bra{f} \otimes \hat{I}\right) \hat{U}_{SM}(s)\right \rangle} \right)^2.
\end{equation}
We are specifically interested in the dependence of this conditional fluctuation on the measurement strength $s$ in the limit of weak interactions. It should be noted that the post-selection has no effect on the initial meter fluctuations at $s = 0$ since the meter and the system are not correlated before the measurement interaction is applied. Likewise, the symmetry of an unbiased meter described by Eqs. (\ref{eq:inverse}), (\ref{eq:Ssymmetry}) and (\ref{eq:Bsymmetry}) guarantees that there is no linear dependence of post-selected meter fluctuations on $s$. The lowest order effect of the post-selected interaction on the meter fluctuations is given by
\begin{equation}
\Delta M^2(s \mid f) = \Delta M^2(-s \mid f).
\end{equation}
The first derivative of $\Delta M^2(s \mid f)$ at $s=0$ is therefore zero. Any changes to the initial meter fluctuations appear only in the second order dependence on $s$, whether they are conditioned by a subsequent measurement or not. 

The complete second derivative of the conditional meter fluctuations $\Delta M^2(s \mid f)$ has many different contributions involving both first and second derivatives of the Heisenberg picture dependence of operators on interaction strength $s$. The main new contribution to the dynamics is the second derivative of the projector $\ket{f}\bra{f}(s)$ in the interaction strength $s$,
\begin{equation}
\label{eq:nlfchange}
\left.\frac{d^2}{ds^2} \ket{f}\bra{f}(s) \right|_{s=0} = - \frac{1}{\hbar^2} \left[\hat{A}, \left[\hat{A}, \ket{f}\bra{f} \right] \right] \otimes \hat{B}^2.
\end{equation}
The sequential commutation relations in this equation describe the second derivative of the dependence of the outcome $\ket{f}\bra{f}$ on a unitary transformation generated by $\hat{A}$. Using the unitary transform defined by Eq.(\ref{eq:Sdynamics}), 
\begin{equation}
\left.\frac{d^2}{d\varphi^2_A} \ket{f}\bra{f}(\varphi_A)\right|_{\varphi_A=0}
= - \frac{1}{\hbar^2} \left[\hat{A}, \left[\hat{A}, \ket{f}\bra{f} \right] \right].
\end{equation}
The commutation relations thus express the effects of the system dynamics on the probability of finding $\ket{f}$ in a subsequent measurement. Eq.(\ref{eq:nlfchange}) shows that these changes in the probability of $\ket{f}$ depend on the magnitude of the squared meter generator $\hat{B}^2$. 

We can now find the second derivative of the conditional meter fluctuations $\Delta M^2(s\mid f)$ by using Eqs.(\ref{eq:lshift}), (\ref{eq:nlshift}), (\ref{eq:lfchange}) and (\ref{eq:nlfchange}). The various contributions to the second derivative are given by
\begin{eqnarray}
\label{eq:cond.fluc.}
\left.\frac{d^2}{ds^2} \Delta M^2(s\mid f) \right|_{s=0} &=& \left(2\braket{\hat{\Gamma}_M}^2 \varepsilon^2_A(f) + 2\Delta \Gamma^2_M \frac{\bra{f} \hat{A} \hat{\rho} \hat{A} \ket{f}}{\bra{f} \hat{\rho} \ket{f}} 
- \braket{\hat{M} \hat{\Theta}_M + \hat{\Theta}_M \hat{M}} \frac{\bra{f} \hat{A} \hat{\rho} \hat{A} \ket{f}}{\bra{f} \hat{\rho} \ket{f}} \right) \nonumber \\
&+&  \left(\frac{1}{2}\braket{\hat{B}^2 \hat{M}^2 + \hat{M}^2 \hat{B}^2} - \braket{\hat{B}^2} \braket{\hat{M}^2} \right) \frac{\frac{d^2}{d \varphi_A^2} \bra{f} \hat{\rho} \ket{f}}{\bra{f} \hat{\rho} \ket{f}},
\end{eqnarray}
where
\begin{equation}
\label{eq:Ozawa}
\varepsilon^2_A(f) = \frac{\bra{f} \hat{A} \hat{\rho} \hat{A} \ket{f}}{\bra{f} \hat{\rho} \ket{f}} - \left(\mathrm{Re} \left(\frac{\bra{f} \hat{A} \hat{\rho} \ket{f}}{\bra{f} \hat{\rho} \ket{f}}\right)\right)^2.
\end{equation}
Eq.(\ref{eq:cond.fluc.}) is the central result of the present analysis. First, we should compare this result to the evolution of meter fluctuations when no subsequent measurement is performed, as shown in Eq.(\ref{eq:decomposition}). The three terms in the first set of brackets can be identified with the corresponding terms in Eq.(\ref{eq:decomposition}), where the expectation value of $\hat{A}$ is replaced by the real part of the weak value and the expectation value of $\hat{A}^2$ is replaced by a symmetric operator expression where the operators $\hat{A}$ are sandwiched between the density operator $\hat{\rho}$ and the outcome $\ket{f}$ of the subsequent measurement. The conditional fluctuations of $\hat{A}$ are represented by $\varepsilon^2_A(f)$, an expression that corresponds to the measurement error originally introduced by Ozawa in \cite{Oza03} and later shown to be experimentally observable by feedback compensation \cite{Hof21, Lem22, Hof23}. The present analysis thus confirms that this expression describes the physical effects of conditional fluctuations of $\hat{A}$. Likewise, the terms associated with fluctuations of the meter response and with meter saturation show that the conditional value of $\hat{A}^2$ is given by the symmetric product with the density operator $\hat{\rho}$ in the middle, consistent with its appearance in the definition of the Ozawa uncertainty $\varepsilon^2_A(f)$, as shown in Eq.(\ref{eq:Ozawa}). It may be worth noting that, for pure states only, the first term in Eq.(\ref{eq:Ozawa}) is equal to the absolute value square of the weak value of $\hat{A}$, such that the Ozawa uncertainty is given by the square of the imaginary part of the weak value. Here, the real part of the weak value represents the best estimate of $\hat{A}$ for the initial state and the post-selected outcome $\ket{f}$ \cite{Hal04}. On the other hand, the imaginary part of the weak value can be interpreted as the optimal estimator for the parameter $\varphi_A$ of the back-action dynamics. As shown in \cite{Hof11}, there is a trade-off relation between the Fisher information of $\varphi_A$ and the estimate of the value of $\hat{A}$, resulting in the above limit of the Ozawa uncertainty. Here, the observed meter fluctuations are most directly related to the fluctuations of the real values of $\hat{A}$ around the best estimate given by the real part of the weak value.

The final term in Eq.(\ref{eq:cond.fluc.}) has no correspondence in Eq. (\ref{eq:decomposition}) and therefore describes an effect exclusively caused by the selection of a subsequent measurement outcome $\ket{f}$. As suggested by the second derivative of the probability $\bra{f}\hat{\rho}\ket{f}$ in $\varphi_A$, this term originates from the dynamics of the system and does not concern physical changes of the meter readout $\hat{M}$. Instead, the reason why the effects of the internal system dynamics appear in the conditional fluctuations $\Delta M^2(s\mid f)$ is the correlation between the magnitudes of the meter generator $\hat{B}^2$ and the meter readout $\hat{M}^2$. Since the amount by which the probability of finding the outcome $\ket{f}$ is changed by the interaction depends on the magnitude of the squared meter generator $\hat{B}^2$, the observation of $\ket{f}$ results in a Bayesian update of the statistics of $\hat{B}^2$. If $\hat{B}^2$ and $\hat{M}^2$ were statistically independent, this Bayesian update would have no effects on the meter fluctuations. However, the meter states defines a correlation between the meter generator and the meter readout given by 
\begin{equation}
K_{MB} = \frac{1}{2}\braket{\hat{B}^2 \hat{M}^2 + \hat{M}^2 \hat{B}^2} - \braket{\hat{B}^2} \braket{\hat{M}^2}.
\end{equation}
According to this correlation, a Bayesian update of $\hat{B}^2$ will result in a corresponding update of $\hat{M}^2$. Effectively, the subsequent measurement of $\ket{f}$ reveals new information about the magnitude of $\hat{M}^2$ before the interaction between the meter and the system, resulting in a change of the readout fluctuations that has nothing to do with the physical meter shift caused by a unitary $\hat{U}(\varphi_B)$ as given by Eq.(\ref{eq:Munitary}). 

The final term in Eq.(\ref{eq:cond.fluc.}) can describe an increase or a decrease in the meter fluctuations, depending on the signs of the second derivative and the sign of the correlation $K_{MB}$. The second derivative in $\varphi_A$ determines whether the probability of $\ket{f}$ increases or deceases as a result of the back-action dynamics. For positive values of the second derivative, finding $\ket{f}$ makes large values of $\hat{B}^2$ more likely. This can result in a reduction of meter fluctuations if the correlation $K_{MB}$ is negative. It should be noted that negative correlations ($K_{MB}<0$) are more likely than positive correlations, since typical meter states will be minimal uncertainty states of $\hat{M}$ and $\hat{B}$, so that any decrease in the expectation value of $\hat{M}^2$ requires an increase in the expectation value of $\hat{B}^2$. For the same reason, it is unlikely that $\hat{M}^2$ and $\hat{B}^2$ are uncorrelated. In most post-selected measurements, the conditional meter fluctuations will therefore include a Bayesian update associated with a negative correlation of $K_{MB}<0$.

\section{Continuous variable meters and Gaussian meter states}
\label{sec:typical}
The theory developed in this paper applies to a wide variety of meter systems. In particular, no assumption is made about the eigenvalues of $\hat{M}$ or about the meter response $\hat{\Gamma}_M$. Since most measurement theories try to achieve a particularly simple mapping between individual meter readouts and the possible values of the physical property $\hat{A}$, it is often assumed that the meter readout is a continuous variable and the initial meter state is a Gaussian. In this section, we show that, despite the simplifications introduced by this intuitive model of measurements, the problem of the correlation between meter readout and meter generator remains. 

In measurement theories with continuous variable, the meter readout $\hat{M}$ is usually identified with the meter position $\hat{x}$ and the meter generator $\hat{B}$ is identified with the conjugate momentum $\hat{p}$. With this assignment of meter properties, the meter response $\hat{\Gamma}_M$ is constant and normalized to one,
\begin{equation}
\hat{\Gamma}_M = \hat{I}.
\end{equation} 
Since the meter response is constant, the associated saturation $\hat{\Theta}_M$ is zero, 
\begin{equation}
\hat{\Theta}_M = 0.
\end{equation} 
The meter response does not depend on the measurement strength $s$ and the Heisenberg equations of motion for the readout $\hat{x}(s)$ can be integrated, resulting in a meter shift $\hat{x}(s)$ that is proportional to the system property $\hat{A}$,
\begin{equation}
\label{eq:xs}
\hat{x}(s) = \hat{I} \otimes \hat{x} + s \hat{A} \otimes \hat{I}.
\end{equation}
It is easy to see why this description of a meter system corresponds to the intuitive idea of a measurement of $\hat{A}$. Independent of the magnitude of the values of $\hat{A}$ and $s$, the change in the meter position $\hat{x}(s)-\hat{x}(0)$ is a direct measure of the physical property $\hat{A}$. The fluctuations of the meter readout can now be determined directly from Eq.(\ref{eq:xs}). Since there are no correlations between the initial fluctuations of the meter and the fluctuations of $\hat{A}$, the measurement interaction always increases the readout fluctuations by adding the fluctuations of $\hat{A}$ to the readout fluctuations $\Delta x^2$ of the initial meter state,
\begin{equation}
\label{eq:decom.x}
\Delta x^2(s) = \Delta x^2 + s^2 \Delta A^2.    
\end{equation}
As expected, this equation satisfies the differential equation for the readout fluctuation in Eq.(\ref{eq:decomposition}) for $\hat{\Theta}_M=0$ and $\Delta \Gamma^2_M=0$. The measurement readout thus reproduces the statistics of the physical property $\hat{A}$ with additional uncorrelated errors given by the independent statistics of the initial meter state. 

A subsequent measurement performed on the system will provide additional information about the actual value of $\hat{A}$, resulting in a change of the conditional fluctuations of $\hat{A}$ corresponding to a Bayesian update of the statistics of $\hat{A}$ based on the subsequent outcome $\ket{f}$. As we have seen in the previous analysis, this updated fluctuation of $\hat{A}$ is given by the Ozawa uncertainty $\varepsilon^2_A(f)$ given by Eq.(\ref{eq:Ozawa}). If the measurement of $\ket{f}$ provided no information about the initial fluctuations of the meter readout, it would be sufficient to replace $\Delta A^2$ with $\varepsilon^2_A(f)$ in Eq.(\ref{eq:decom.x}) to obtain the conditional readout fluctuations $\Delta x^2(s\mid f)$. However, Eq.(\ref{eq:cond.fluc.}) shows that this is only correct when there are no correlations between $\hat{x}^2$ and $\hat{p}^2$ ($K_{xp}=0$). The dependence of the post-selected meter fluctuation on $s$ cannot be integrated and we can only determine the weak interaction limit using the second derivative,
\begin{eqnarray}
\label{eq:cond.x}
\left.\frac{d^2}{ds^2} \Delta x^2(s | f) \right|_{s=0} = 2 \varepsilon^2_A(f) + K_{xp} \frac{\frac{d^2}{d\varphi_A^2}  \bra{f} \hat{\rho} \ket{f}}{\bra{f} \hat{\rho} \ket{f}},
\end{eqnarray}
where
\begin{equation}
K_{xp} = \frac{1}{2}\braket{\hat{p}^2 \hat{x}^2 + \hat{x}^2 \hat{p}^2} - \braket{\hat{p}^2} \braket{\hat{x}^2}.
\end{equation}
In general, the subsequent measurement result $\ket{f}$ provides information about the magnitude of the initial fluctuations of the meter readout via the correlations $K_{xp}$, which is typically negative for states that are close to the uncertainty limit of $\Delta x \Delta p \geq \hbar/2$. The minimal uncertainty states of the meter are Gaussian states with a standard deviation of $\sigma_x^2=\Delta x^2$,
\begin{equation}
\braket{x|\phi} = \frac{1}{(2 \pi \sigma^2_x)^{1/4}} \exp\left(- \frac{x^2}{4\sigma^2_x}\right).\end{equation}
To understand why these states exhibit a correlation between $\hat{x}^2$ and $\hat{p}^2$, it is useful to remember that they can be interpreted as ground states of a harmonic oscillator with an eigenvalue of $1/2$ for a specific linear combination of $\hat{p}^2$ and $\hat{x}^2$. This precise eigenvalue indicates that the fluctuations of $\hat{p}^2$ and $\hat{x}^2$ are anti-correlated. Mathematically, the correlation $K_{xp}$ for all Gaussian states is given by 
\begin{equation}
K_{xp} = - \frac{\hbar^2}{2}.
\end{equation}
The correlation between $\hat{p}^2$ and $\hat{x}^2$ is a quantum effect determined entirely by the constant $\hbar$. This means that the effects of the back-action dynamics of the probability of finding $\ket{f}$ on the conditional readout fluctuations $\Delta x^2(s|f)$ will be the same for all Gaussian states, with a constant ratio of $-\hbar^2/2$ relating the dynamics generated by $\hat{A}$ to the apparent statistics of $\hat{A}$. This fact may give rise to serious misunderstandings because the quantum formalism already relates the second derivatives of probabilities to expressions that look like second order statistical moments of the generator $\hat{A}$. Specifically, the term that appears in the conditional readout fluctuations $\Delta x^2(s|f)$ can be written as 
\begin{equation}
\label{eq:dyn.sta.}
\frac{\frac{d^2}{d\varphi_A^2} \bra{f} \hat{\rho} \ket{f}}{\bra{f} \hat{\rho} \ket{f}} = - \frac{2}{\hbar^2} \left(\mathrm{Re} \left(\frac{\bra{f} \hat{A}^2 \hat{\rho} \ket{f}}{\bra{f} \hat{\rho} \ket{f}}\right) - \frac{\bra{f} \hat{A} \hat{\rho} \hat{A} \ket{f}}{\bra{f} \hat{\rho} \ket{f}}\right).
\end{equation}
Multiplication with $K_{xp}$ results in an expression that looks like a description of conditional statistics. The first term in the brackets corresponds to the real part of the weak value of $\hat{A}^2$ and the second term corresponds to the expression of the conditional value of $\hat{A}^2$ found in Eq.(\ref{eq:cond.fluc.}) and in the conditional fluctuation $\varepsilon_A^2(f)$ given by Eq.(\ref{eq:Ozawa}). Effectively, the dynamics appears to replace expressions of $\hat{A}^2$ where the density operator $\hat{\rho}$ is sandwiched between the operators $\hat{A}$ with expressions for the real part of the weak value of $\hat{A}^2$ where the operators are both on the same side of $\hat{\rho}$. When Gaussian meter states are used, the effects of the conditional fluctuation $\varepsilon_A^2(f)$ combines with this dynamical effect to define an increase in the conditional readout fluctuation that seems to depend only on the post-selected quantum statistics of $\hat{A}$, 
\begin{equation}
\label{eq:only.s}
\left.\frac{d^2}{ds^2} \Delta x^2(s\mid f) \right|_{s=0}  = \mathrm{Re} \left(\frac{\bra{f} \hat{A}^2 \hat{\rho} \ket{f}}{\bra{f} \hat{\rho} \ket{f}}\right) + \frac{\bra{f} \hat{A} \hat{\rho} \hat{A} \ket{f}}{\bra{f} \hat{\rho} \ket{f}} - 2\left(\mathrm{Re} \left(\frac{\bra{f} \hat{A} \hat{\rho} \ket{f}}{\bra{f} \hat{\rho} \ket{f}}\right)\right)^2.
\end{equation}
For mixed states $\hat{\rho}$, it is apparent that the first two terms describe second order statistical moments of $\hat{A}$. The right hand side of the equation could then be interpreted as an average of two different definition of conditional fluctuations, one given by $\varepsilon_A^2(f)$ and based on the theory of Ozawa, and one given by the real parts of the weak values of $\hat{A}^2$ and of $\hat{A}$. In general, the effect of the Bayesian update of the initial meter fluctuations based on the outcome $\ket{f}$ appears to replace half of the conditional uncertainty of $\varepsilon_A^2(f)$ with a description of conditional uncertainties based only on weak values. However, it should be remembered that the weak value of $\hat{A}^2$ is never observed directly and only appears in the equations because of the dependence of the probability of $\ket{f}$ on the dynamics. 

The situation can get even more confusing when the system is initially in a pure state $\ket{\psi}$. In that case, the Ozawa uncertainty $\varepsilon_A^2(f)$ is given by the square of the imaginary part of the weak value, suggesting a similarity of this fluctuation with the square of the real part of the weak value that is not justified by the physics of the measurement interaction. This result has been reported in \cite{Oga21}, where it is identified as a ``weak variance'' of the target observable $\hat{A}$. This weak variance now appears in our analysis of the second derivative of the conditional readout fluctuation as
\begin{eqnarray}
\label{eq:weakvariance}
\left.\frac{d^2}{ds^2} \Delta x^2(s\mid f) \right|_{s=0} = \mathrm{Re}\left(\frac{\bra{f} \hat{A}^2 \ket{\psi}}{\braket{f|\psi}} - \left(\frac{\bra{f} \hat{A} \ket{\psi}}{\braket{f|\psi}}\right)^2\right).
\end{eqnarray}
The concept of a ``weak variance'' is based on the formal analogy with the mathematical definition of a variance, where averages are replaced with weak values. Interestingly, the complex weak value is squared before the real part is selected, so that imaginary parts of the weal value of $\hat{A}$ increase the ``weak variance''. As we have seen above, this addition of the squared imaginary part of the weak value is actually associated with the Ozawa uncertainty $\varepsilon_A^2(f)$ for pure states. 

The interpretation of the right term of Eq.(\ref{eq:weakvariance}) as a ``weak variance'' given in \cite{Oga21} has three problems. First, the ``weak variance'' only describes the conditional meter fluctuations if the initial system state is a pure state $\ket{\psi}$. As we have shown in Eq.(\ref{eq:only.s}), the mixed state case does not include a squared imaginary part of the weak value. Second, the increase in the conditional readout fluctuations caused by the ``weak variance'' is only half of the increase caused by a corresponding uncertainty without any post-selection. It seems that this discrepancy was overlooked in \cite{Oga21}. The analogy between the increase in meter fluctuations caused by fluctuations of $\hat{A}$ and the increase caused by the ``weak variance'' is therefore not valid. Third, we have shown that the physical explanation of the appearance of the weak value of $\hat{A}^2$ in the ``weak variance'' originates from the way in which the back-action changes the probability of obtaining the subsequent outcome $\ket{f}$. In the case of conditional meter readouts, we can trace the appearance of weak values of $\hat{A}^2$ to a Bayesian update of the initial meter fluctuations $\Delta x^2$ based in the increase or decrease of the likelihood of $\ket{f}$ for large or small fluctuations of the meter generator $\hat{p}$. As correctly observed by Ogawa et al. \cite{Oga21}, the meter fluctuations are squeezed as information is obtained about $\langle \hat{p}^2 \rangle$ and $\langle \hat{x}^2 \rangle$ changes in the opposite direction to maintain the uncertainty relations. Importantly, none of these changes originate from any meter dynamics. Instead, information about the conditional back-action dynamics in the system modifies the information we have about the meter state.

\section{Can conditional system fluctuations be observed in the meter readout?}
\label{sec:discussion}
As discussed in the previous section, the ``weak variance'' defined by Ogawa et.al. \cite{Oga21} is actually a linear combination of the conditional fluctuations of the system property $\hat{A}$ described by the Ozawa uncertainty $\varepsilon^2_A(f)$ and a second term representing a Bayesian update of the meter statistics based on the second derivative of the probability $\bra{f}\hat{\rho}\ket{f}$ in $\varphi_A$. The reason why Ogawa et. al. define the ``weak variance'' is that it can be obtained directly from the meter readout statistics by evaluating the conditional fluctuations $\Delta M^2(s\mid f)$. It therefore has some merit as an experimentally accessible quantity, even if there are some problems with its physical meaning. In the following, we consider the circumstances that would allow us to distinguish the different contributions to the meter readout fluctuations in an experiment, especially with regard to a distinction between the Ozawa uncertainties $\varepsilon^2_A(f)$ and the back-action dynamics expressed by the second derivative of the post-selection probability. 

To understand the problem from the viewpoint of conditional fluctuations of $\hat{A}$, it is useful to express the Ozawa uncertainty and the back-action dynamics using the quantum statistical expressions of $\hat{A}$ given by Eqs.(\ref{eq:Ozawa}) and (\ref{eq:dyn.sta.}). We can then obtain a consistent description of the conditional meter fluctuations given by Eq.(\ref{eq:cond.fluc.}) in terms of quantum statistical moments of the system property $\hat{A}$ conditioned by the subsequent outcome $\ket{f}$, 
\begin{eqnarray}
\label{eq:experimentaldata}
\left.\frac{d^2}{ds^2} \Delta M^2(s\mid f) \right|_{s=0} &=& 2\braket{\hat{\Gamma}_M}^2 \left(\frac{\bra{f} \hat{A} \hat{\rho} \hat{A} \ket{f}}{\bra{f} \hat{\rho} \ket{f}} - \left(\mathrm{Re} \left(\frac{\bra{f} \hat{A} \hat{\rho} \ket{f}}{\bra{f} \hat{\rho} \ket{f}}\right)\right)^2\right) \nonumber \\
&+& \left(2\Delta \Gamma^2_M - \braket{\hat{M} \hat{\Theta}_M + \hat{\Theta}_M \hat{M}}\right) \frac{\bra{f} \hat{A} \hat{\rho} \hat{A} \ket{f}}{\bra{f} \hat{\rho} \ket{f}}  \nonumber \\
&-& \frac{2}{\hbar^2}  K_{MB} \left(\mathrm{Re} \left(\frac{\bra{f} \hat{A}^2 \hat{\rho} \ket{f}}{\bra{f} \hat{\rho} \ket{f}}\right) - \frac{\bra{f} \hat{A} \hat{\rho} \hat{A} \ket{f}}{\bra{f} \hat{\rho} \ket{f}}\right).
\end{eqnarray}
Each of the three terms on the left hand side of this equation expresses very different physics. Nevertheless each term includes a contribution from the conditional second order moment of $\hat{A}$ where the density operator $\hat{\rho}$ is sandwiched between the two operators $\hat{A}$. Eq.(\ref{eq:experimentaldata}) thus highlights the difficulty of distinguishing between the different possible origins of these contributions. 

The previous analysis showed that the fluctuations of $\hat{A}$ conditioned by a subsequent measurement outcome $\ket{f}$ are described by the Ozawa uncertainty $\varepsilon_A^2(f)$ that appears where the initial state uncertainty $\Delta A^2$ appears in the second derivative of meter noise without post-selection (Eq.(\ref{eq:decomposition})). This conditional fluctuation of $\hat{A}$ is the origin of the first term in Eq.(\ref{eq:experimentaldata}). The second term also corresponds to terms in Eq.(\ref{eq:decomposition}) and is associated with fluctuations of the meter response and with meter saturation. The appearance of the sandwiched operator ordering $\hat{A}\hat{\rho}\hat{A}$ indicates that this term is the valid representation of the second order statistical moment of $\hat{A}$ in the system-meter interaction. It is therefore possible to interpret the meter fluctuation and saturation in terms of genuine conditional statistics of $\hat{A}$. This is different for the third and final term in Eq.(\ref{eq:experimentaldata}). As we have seen in the previous discussion, this term describes a Bayesian update of the meter readout fluctuations based on the sensitivity of the post-selection probability of $\ket{f}$ on the back-action dynamics generated by $\hat{A}$ in the system. It is quite remarkable that quantum mechanics expresses the dynamics of this post-selection probability in a form that looks like a statistical fluctuation of the generator of the dynamics $\hat{A}$. To highlight this potential source of confusion, we will define the expression that makes the dynamic response appear as a quantum fluctuation of its generator as the dynamic pseudovariance of $\hat{A}$,
\begin{equation}
\label{eq:Vdyn}
V_{\mathrm{dyn.}} := - \frac{\hbar^2}{2} \frac{\frac{d^2}{d\varphi_A^2}\bra{f} \hat{\rho} \ket{f}}{\bra{f} \hat{\rho} \ket{f}}.
\end{equation}
The dynamic pseudovariance is defined in terms of the dynamics generated by $\hat{A}$, but it obtains the characteristics of a variance through the factor of $-\hbar^2/2$. Quantum mechanics naturally establishes this relation between dynamics and statistics through the role of commutation relations in the definition of the dynamics. In general, pseudovariances are as likely to be negative as they are to be positive. However, in the case of pure states $\ket{\psi}$, the definition of the pseudovariance results in a difference between the weak value of $\hat{A}^2$ and the absolute square of the weak value, corresponding to another possible definition of weak variances, slightly different from the one suggested in \cite{Oga21} and Eq.(\ref{eq:weakvariance}),
\begin{equation}
\label{eq:pureVdyn}
V_{\mathrm{dyn.}} = \mathrm{Re}\left(\frac{\bra{f} \hat{A}^2 \ket{\psi}}{\braket{f|\psi}} \right) - \left| \frac{\bra{f} \hat{A} \ket{\psi}}{\braket{f|\psi}}\right|^2.
\end{equation}
For pure states only, the dynamic pseudovariance evaluates the difference between the real weak value of $\hat{A}^2$ and the absolute value squared of the weak value of $\hat{A}$. In general, the weak value of $\hat{A}^2$ can take a wide range of values, both positive and negative. However, it may be worth noting that there are cases where the weak value of $\hat{A}^2$ is independent of the initial state and of post-selections. If $\hat{A}$ only has eigenvalues of $+1$ or $-1$, the square is equal to the identity, $\hat{A}^2=\hat{I}$, and the weak value of $\hat{A}^2$ will be always be equal to one. This is typically the case in two-level systems, and it entered into the analysis given in \cite{Ber15}, making it difficult to identify the role of the weak value played in that work. On the other hand, the sandwiched operator expression of the square of $\hat{A}$ is not limited to values of one by eigenvalues of $\pm 1$, so the evaluation of the conditional meter fluctuations can be used to obtain the correct value of the conditional fluctuations $\varepsilon_A^2(f)$ based on an appropriate analysis of the different contributions to the experimentally observed result. 

If $\hat{A}^2$ is not equal to the identity, a separate measurement is needed to obtain either the weak value of $\hat{A}^2$ or the dynamic pseudovariance $V_{\mathrm{dyn.}}$. In principle, $V_{\mathrm{dyn.}}$ can be determined from its definition given in Eq.(\ref{eq:Vdyn}) by determining the dependence of the post-selection probability of $\ket{f}$ on small changes in the parameter $\varphi_A=s B_b$. For a meter state with a given statistical variance $\langle \hat{B}^2 \rangle$ of the meter generator $\hat{B}$, such data can be obtained by leaving $s$ unchanged, using the statistical variations of $\hat{B}$ to generate a sufficient data set of different values of $\varphi_A=s B_b$. The second derivative of $\bra{f} \hat{\rho} \ket{f}$ in $\varphi_A$ can then be determined from the joint probabilities of $\ket{f}$ and $B_b$ observed at a fixed value of $s$. It might be worth noting that this is similar to the evaluation of the imaginary part of the weak value from the conditional average of $B_b$, indicating that the imaginary weak value also describes a Bayesian update of the meter system due to the dependence of the post-selection probability on $\varphi_A=s B_b$. In the present context, the essential point is that the system dynamics needed for an experimental evaluation of the dynamic pseudovariance is already provided by the measurement interaction, and it can be controlled with sufficient precision by determining the value of $B_b$ after the interaction.

For the sake of simplicity, it may also be interesting to construct a meter state that has a correlation of $K_{MB}=0$. Eq.(\ref{eq:experimentaldata}) shows that the conditional fluctuations $\varepsilon_A^2(f)$ can then be determined from the conditional readout fluctuations and the weak value of $\hat{A}$ determined by the conditional meter shift $\langle \hat{M}(s|f)\rangle$. However, most meter states have a non-vanishing correlation $K_{MB}$, as discussed above for Gaussian states. The most simple case of $K_{MB}=0$ is obtained when either $\hat{M}$ or $\hat{B}$ only have eigenvalues of $+1$ or $-1$. Although this condition might seem overly restrictive, it is usually satisfied when the meter is a single qubit existing in a two dimensional Hilbert space. However, there is a problem with such a meter system. Since $\hat{M}^2=\hat{B}^2=1$, the meter readout fluctuations can be determined directly from the conditional meter shift $\langle \hat{M}(s|f)\rangle$, and this shift is necessarily independent of the intrinsic fluctuations of $\hat{A}$. This observation can be confirmed by calculating the effects of meter fluctuations and meter saturation. Typically, the initial meter state will be an eigenstate of $\hat{\Gamma}_M$ with an eigenvalue of one $(\langle \hat{\Gamma}_M\rangle=1)$, and the saturation will be given by $\Theta_M=\hat{M}$. The sandwiched expression of the square of $\hat{A}$ in the Ozawa uncertainty will then be cancelled exactly by the saturation term associated with the limited eigenvalues of the meter readout. It is therefore impossible to obtain any information about second order statistical moments of $\hat{A}$ from a two-level meter. 

It is possible to construct meter states that have $K_{MB}=0$ in continuous variable Hilbert spaces, so that the increase in meter fluctuations can serve as a direct measure of the conditional fluctuations of $\hat{A}$. However, these meter states will be very different from the minimal uncertainty states used to optimize the uncertainty trade-off between measurement resolution and disturbance \cite{Mat21}, and it may be difficult to quantify the change of the readout fluctuations for the expected non-Gaussian statistics. In general, it is a highly non-trivial task to determine the conditional fluctuations of $\hat{A}$ from the data obtained in a meter readout. In most cases, additional statistical information will be needed. The problems caused by the appearance of the Bayesian update of the meter state as a dynamical pseudovariance of $\hat{A}$ means that the simple assumption that any change in the readout fluctuations originates from the corresponding conditional fluctuations of $\hat{A}$ is unfortunately not valid.

\section{Conclusion}
\label{sec:Conclusion}
We have investigated the physical origin of the changes in the meter readout fluctuations observed in weak measurement interactions. A variety of effects has been identified. Even without post-selection of a subsequent measurement result, non-linearities in the meter dynamics can introduce changes that do not reflect the intrinsic uncertainties of the target observable. In particular, we find that meter saturation effects can reduce the readout fluctuations for large values of the target observable. The simple picture of an addition of intrinsic fluctuations of the target observable to the initial readout fluctuations thus already breaks down for non-linear meter shifts. When post-selection is added to the picture, the additional information about the target observable $\hat{A}$ results in a Bayesian update of the statistics. Our analysis of the conditional changes in the readout statistics confirms that the updated expectation value of $\hat{A}$ is given by the weak value. Likewise, the updated variance of $\hat{A}$ is given by the Ozawa uncertainties, indicating that Ozawa uncertainties are the correct representation of the fluctuations of $\hat{A}$ under the condition set by the outcome of a subsequent measurement. However, we also find an additional term that appears in the meter readout fluctuations only when post-selection is applied in a subsequent measurement. The analysis shows that this term relates to information obtained in the post-selection measurement about the magnitude of the back-action dynamics caused by the fluctuations of the meter generator $\hat{B}$. As a result of this Bayesian update of the fluctuations of the meter generator $\hat{B}$, the fluctuations of the meter readout $\hat{M}$ will also be updated if there is a correlation $K_{MB}$ between the fluctuations of the squared generator $\hat{B}^2$ and the squared readout $\hat{M}^2$. The subsequent measurement therefore not only updates the fluctuations of the system property $\hat{A}$, but also adds a specific multiple of the second derivative of the probability $\bra{f}\hat{\rho}\ket{f}$ in the parameter $\varphi_A$ of the unitary dynamics generated by $\hat{A}$ to the conditional meter fluctuations, reflecting the information about the initial fluctuations of the readout observable $\hat{M}$ obtained by the measurement of $\ket{f}$ in the system. In most cases, meter states are minimal uncertainty states of $\hat{B}$ and $\hat{M}$ characterized by negative correlations $K_{MB}<0$. It is therefore difficult to avoid an effect of the information gained about the initial meter fluctuations on the readout fluctuations. In particular, a negative correlation of $K_{MB}=-\hbar^2/2$ is obtained for Gaussian meter states, explaining the reduction of readout fluctuations reported in \cite{Ber15}. Effectively, the information provided by the post-selection can improve the conditional precision of a weak measurement by indicating that the initial meter fluctuation was lower than average in the post-selected sub-ensemble. 

Our analysis can distinguish between contributions associated with the actual fluctuations in the value of $\hat{A}$, contributions associated with meter non-linearities determined by the magnitude of $\hat{A}^2$, and contributions associated with the information obtained in a subsequent measurement of $\ket{f}$ determined by the dependence of the probability of $\ket{f}$ on the dynamics generated by $\hat{A}$. Our results show that the precise contributions associated with post-selection depend on the initial meter state. This is particularly important with respect to the result previously reported by Ogawa et al. \cite{Oga21}, where the total effect on the meter readout fluctuations was identified with a ``weak variance'' given by the real part of the weak value of $\hat{A}^2$ minus the square of the complex weak value of $\hat{A}$. Our result now show that this effect is only observed with Gaussian meter states, since the ``weak variance'' is actually a combination of Ozawa uncertainties representing the fluctuations of $\hat{A}$ conditioned by $\ket{f}$ and dynamic pseudovariance associated with a negative correlation of $K_{MB}= - \hbar^2/2$ for Gaussian meter states. The experiment in \cite{Oga21} should therefore be repeated using a non-Gaussian meter state with a different correlation $K_{MB}$. Our results predict that this will result in the observation of a different variance of $\hat{A}$ in the readout fluctuations. It may then be possible to distinguish the conditional uncertainty of $\hat{A}$ from the dynamical pseudovariance in an experiment. A precise evaluation of the date provided in \cite{Oga21} also confirms that the ``weak variance'' adds only half as much to the readout fluctuations as a conventional fluctuation of $\hat{A}$. The correct explanation of this effect is that the "weak variance" is equal to a sum of the Ozawa uncertainty describing the conditional fluctuations of $\hat{A}$ and a dynamic pseudovariance that effectively replaces half of the quadratic term in the Ozawa uncertainties with the weak value of $\hat{A}^2$. Ogawa et al. \cite{Oga21} suggested that the ``weak variance'' might be linked to the statistics of quasi-probabilities determined in weak measurements \cite{Lun12,Hof12}. It is therefore interesting to note that their result originates from a combination of Ozawa uncertainties with the dynamics generated by the same observable $\hat{A}$. This seems to support the idea that the quasi-probabilities obtained in weak measurements are an expression of reversible dynamics in quantum mechanics \cite{Hof12,Hof15}. Nevertheless our analysis shows that the fluctuations of a physical quantity defined by pre- and post-selection are given by Ozawa's definition of uncertainties and are therefore always positive. 

Our results also show that the reduction of the meter fluctuations reported in \cite{Par11, Par13, Ber15, Mat17} originates from a Bayesian update of the initial meter fluctuations based on a combination of the observed back-action dynamics with a negative correlation of $K_{MB}<0$. It should be noted that this problem is very different from a parameter estimation problem, even though it involves very similar statistical elements. In particular, weak values have been widely used in quantum metrology to enhance the precision of parameter estimation \cite{Bru10, Xu13, Mag14, Zhn15, Pan15, Har17}. In the present context, it is possible to apply these methods to the optimization of the meter readout $\hat{M}$, where the imaginary weak values of the meter generator $\hat{B}$ would appear as eigenvalues of $\hat{M}$ \cite{Hof11, Dre12, Hof12}. 
However, these methods have no impact on the initial meter noise, which determines the intrinsic uncertainty limit of the measurement. The conditional reduction in meter noise by post-selection makes use of the additional information about the initial meter noise obtained because the measurement back-action effectively serves as a measurement of the meter. This effect identifies fluctuations in meter sensitivity associated with the quantum uncertainty of the initial meter state by relating them to the back-action dynamics of the system. Post-selection can enhance the sensitivity of the meter by focusing on fluctuations where the meter noise was below average. It is interesting to note that the back-action of a measurement thus provides the means of increasing the sensitivity of a measurement by post-processing of the meter information transferred to the system that is being measured, illustrating the importance of a more detailed analysis of the entanglement between the system and the meter generated by the measurement interaction.

Primarily, our analysis shows that the intrinsic fluctuations of weak values determined by post-selections are given by the Ozawa uncertainties in Eq.(\ref{eq:Ozawa}). In the case of pure state inputs, these uncertainties are given by the square of the imaginary part of the weak value, indicating that the weak values of $\hat{A}$ represent the precise value of the post-selected observable if their imaginary parts are zero \cite{Hof21,Lem22,Vai17,Dzi19}. It is difficult to determine these intrinsic fluctuations experimentally because the change in the meter fluctuations caused by the measurement interaction also includes the effects of non-linear meter dynamics and the update of meter fluctuations associated with post-selection. Our results thus highlight the importance of considering the precise physics that shape the quantum statistics of a meter readout in a quantum measurement.

\section*{acknowledgment}
This work was supported by JST, the establishment of university fellowships towards the creation of science technology innovation, Grant Number JPMJFS2129.

\end{document}